\documentclass[twocolumn,prb,showpacs]{revtex4}
\usepackage{graphicx}
\usepackage{dcolumn}
\usepackage{bm}
\begin{document}
\title{Phase diagram of the one-dimensional Hubbard model \\ 
with next-nearest-neighbor hopping}
\author{S.~Nishimoto,$^{1}$ K.~Sano,$^{2}$ and Y.~Ohta$^{3}$}
\affiliation{$^{1}$Max-Planck-Institut f\"ur Physik komplexer Systeme, 
D-01187 Dresden, Germany\\
$^{2}$Department of Physics Engineering, Mie University, 
Tsu 514-8507, Japan\\
$^{3}$Department of Physics, Chiba University, Chiba 263-8522, Japan}
\date{\today}
\begin{abstract}
We study the one-dimensional Hubbard model with nearest-neighbor and 
next-nearest-neighbor hopping integrals by using the density-matrix 
renormalization group (DMRG) method and Hartree-Fock approximation.  
Based on the calculated results for the spin gap, total-spin quantum number, 
and Tomonaga-Luttinger-liquid parameter, we determine the ground-state 
phase diagram of the model in the entire filling and wide parameter region.  
We show that, in contrast to the weak-coupling regime where a spin-gapped 
liquid phase is predicted in the region with four Fermi points, the spin gap 
vanishes in a substantial region in the strong-coupling regime.  
It is remarkable that a large variety of phases, such as the paramagnetic 
metallic phase, spin-gapped liquid phase, singlet and triplet superconducting 
phases, and fully polarized ferromagnetic phase, appear in such a simple 
model in the strong-coupling regime.  
\end{abstract}
\maketitle

\section{INTRODUCTION}

For several decades, quasi-one-dimensional (1D) materials have 
been one of the major subjects of research in the field of condensed 
matter physics.~\cite{ishiguro90,kiess92,kishida00} 
A standard description of such materials is the 1D Hubbard 
model.~\cite{Gut63,Hub63,Kan63} 
The simplest case with the cosine dispersion (nearest-neighbor 
hopping only) was solved exactly by Lieb and Wu via the Bethe 
ansatz.~\cite{Lie68}  The low-lying excitations were also 
understood well as the Tomonaga-Luttinger liquid (TLL),~\cite{Hal80} 
where the renormalization group technique and bosonization method 
have been used.~\cite{Bal96}  
However, modifications of the 1D Hubbard model are often 
required for realistic descriptions of the materials.  
In general, such modifications (even if they are small) make 
the analyses much more difficult since the correlation effects 
are strong in low-dimensional systems.  
Thus, even in the 1D systems, our knowledge is still far from 
being complete.

One of the typical modifications is to add a next-nearest-neighbor 
hopping term in the Hamiltonian, which brings a sort of 
frustration to the spin degrees of freedom of the system as 
well as some coupling between spin and charge degrees of freedom.  
In the past, this model has been extensively studied and some 
distinctive features, which are absent in the simple 1D Hubbard 
model, have been found.  
At half filling, the system has three phases: 
one is a Mott insulating phase with $2k_{\rm F}$ spin-density-wave 
(SDW) correlation (which occurs when the spin frustration is 
small); the others are a spin-gapped insulating phase with 
incommensurate spiral correlation and a spin-gapped metallic 
phase for sufficiently large spin frustration.~\cite{Kur97,Dau00,Tor03,Jap06}  
Away from half filling, the existence of ferromagnetism has 
been shown analytically in some limiting cases for infinite 
strength of the coupling,~\cite{Mat74,Sig92,Mue93} which has been 
confirmed numerically for finite but large enough strength of the 
coupling.~\cite{Pie96,Dau971,Dau972}  
Also, it has been pointed out that, although a weak-coupling 
analysis leads to only a spin-singlet superconducting phase 
with finite spin gap,~\cite{Fab96} previous density-matrix 
renormalization group (DMRG) studies suggest that the spin gap 
vanishes for large enough coupling strengths when the 
next-nearest-neighbor hopping is positive and 
large.~\cite{Ari98,Dau00}  
Moreover, a spin-triplet superconducting phase has been shown 
to exist at quarter filling.~\cite{Oht05}  
As just described, the present system has many phases unparalleled 
in other 1D strongly-correlated electron systems; i.e., our 
modified 1D model can involve a variety of physical phenomena.  
In particular, to detect their phases is of particular interest 
in the light of recent proposals to realize a Hubbard model of 
fermions on an optical lattice.~\cite{Hof02}  

There are some relevant materials to the 1D Hubbard model with 
next-nearest-neighbor hopping.  One is the quasi-1D organic conductor 
(TMTSF)$_2$X [X=PF$_6$, ClO$_4$], the so-called Bechgaard 
salt.~\cite{Jer80,Bec81}  This material exhibits a rich phase diagram upon 
variation of the pressure and temperature.  At low temperatures, 
the phase changes in the order as the spin-Peierls insulator, 
antiferromagnetic insulator, spin-density-wave (SDW) insulator, 
superconductor, and paramagnetic metal, with increasing pressure.  
So far, experimental evidence that the superconducting 
state is in the spin-triplet channel has been piled up.~\cite{Lee02,Lee06}  
Theoretically, it has been proposed that a triangular lattice formed 
by the hopping integrals makes the ferromagnetic ring-exchange 
mechanism relevant, which in turn leads to the spin-triplet 
superconductivity.~\cite{Oht06} 

Another relating system is a newly synthesized copper-oxide compound 
Pr$_2$Ba$_4$Cu$_7$O$_{15-\delta}$.~\cite{Mat04}  This material 
consists of both the single CuO chains (as in PrBa$_2$Cu$_3$O$_7$) 
and double CuO chains (as in PrBa$_2$Cu$_4$O$_8$), and those 
chains are separated by insulating CuO$_2$ plains.  It has been 
reported that the double chains turn into a superconducting state 
below $T_{\rm c} \sim 10$ K.~\cite{Sas06}  So far, some numerical 
studies have been carried out; on the basis of the TLL theory, 
a weak-coupling phase diagram has been obtained in the $d$-$p$ 
double chain model.~\cite{San05,San07}  Also, in a reduced single-band 
double chain model, it has been suggested that the superconducting 
gap has an extended $s$-wave-like form, which does not contradict 
with the experimental results.~\cite{Nak07}  
Relevance of charge fluctuations has also been discussed.~\cite{Ama,Nis03}

In this paper, we study the 1D Hubbard model with the next-nearest-neighbor 
hopping.  We calculate the TLL parameters, spin gap, and total-spin 
quantum number by using the DMRG method and Hartree-Fock (HF) 
approximation.  Based on the results, a detailed phase diagram as 
a function of band filling and hopping integrals is determined 
in both weak-coupling and strong-coupling regimes.  
Surprisingly, the phase diagram of the model contains a large number 
of distinct phases in the strong-coupling regime although the model 
is quite simple.  We hope that the present investigation will contribute 
to better understanding of the 1D strongly-correlated 
electron systems.  

This paper is organized as follows.  In Sec.~II, we define the 
1D Hubbard model with the next-nearest-neighbor hopping and study 
the model in the noninteracting case.  
In Sec.~III, we discuss how the Hartree-Fock approximation and 
DMRG method are used to calculate the TLL parameter.  
In Sec.~IV, we present the calculated results and obtain the phase 
diagrams of the model based on the numerical results.  
Section V contains summary and conclusions.

\section{MODEL}

\begin{figure}[t]
\includegraphics[width=5.3cm]{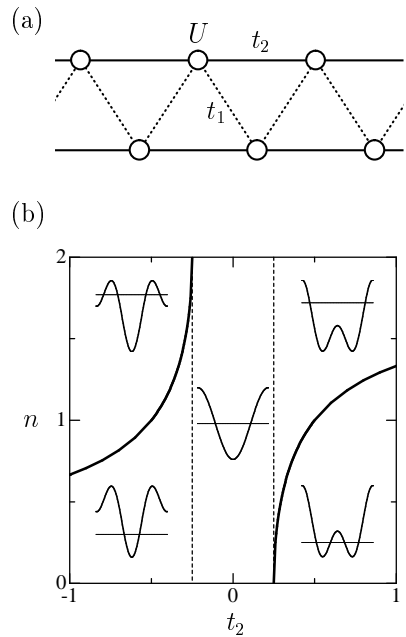}
\caption{(a) Schematic representation of the lattice structure 
of the 1D Hubbard model with the next-nearest-neighbor hopping 
and (b) its $U=0$ phase diagram, where the thick solid curves 
separate the two regimes.  
The inset shows the qualitative behavior of the band dispersion 
$\varepsilon_k$, where the Fermi level is indicated 
by the thin line.}
\label{fig1}
\end{figure}

We consider the 1D Hubbard model with the next-nearest-neighbor 
hopping, which is defined by the Hamiltonian 
\begin{eqnarray}
\nonumber
H&=&t_1\sum_{i,\sigma}(c^\dagger_{i+1 \sigma}c_{i \sigma}+{\rm H.c.}) 
+t_2\sum_{i,\sigma}(c^\dagger_{i+2 \sigma}c_{i \sigma}+{\rm H.c.}) \\
~~~~~&+&U\sum_in_{i\uparrow}n_{i\downarrow}, 
\label{hamiltonian}
\end{eqnarray}
where $c_{i \sigma}^\dagger$ ($c_{i \sigma}$) is the creation 
(annihilation) operator of an electron with spin $\sigma$ at site $i$, 
and $n_{i\sigma}=c_{i \sigma}^\dagger c_{i \sigma}$ is the 
number operator.  $t_1(>0)$ and $t_2$ are the nearest-neighbor 
and next-nearest-neighbor hopping integrals, respectively, and 
$U$ is the on-site Coulomb interaction [see Fig.~\ref{fig1}(a)].  

The dispersion relation is given by
\begin{equation}
\varepsilon_k = 2t_1 \cos ka + 2t_2 \cos 2ka,
\label{dispersion}
\end{equation}
where $k$ is the wave number and $a$ is the lattice constant; 
we set $a=1$ hereafter.  The bare band width is estimated as 
$W=2t_1+4|t_2|+\frac{t_1^2}{4|t_2|}$ for 
$|t_2/t_1| \ge 1/2$ and as $W=4t_1$ for $|t_2/t_1|<1/2$.  
The ground-state phase diagram in the noninteracting case 
($U=0$) is shown in Fig.~\ref{fig1}(b).  
For $|t_2/t_1| < \cos^2[(2-n)\pi/2]/\sin^2[(2-n)\pi]$ 
($n$ is the band filling), the system has two Fermi points 
and the physical properties at low energies are qualitatively 
the same as a system with $t_2=0$.  On the other hand, 
for $|t_2/t_1| > \cos^2[(2-n)\pi/2]/\sin^2[(2-n)\pi]$, 
there are two branches, namely four Fermi momenta 
$\pm k_{\rm F1}$ and $\pm k_{\rm F2}$ ($|k_{\rm F2}|>|k_{\rm F1}|$).  
In this case, as discussed in Sec.~IV.A, the Fermi surface 
can be mapped to that of a two-leg Hubbard ladder model at 
weak coupling.  We designate the critical boundary at which 
the Fermi surface splits into four points as the Fermi-point 
(FP) boundary and the FP boundary is characterized by a 
condition $k_{F1}=0$ (or $k_{F2}=\pi$).  Hence, the model 
(\ref{hamiltonian}) has to be dealt with as a two band system 
within the TLL theory.  Note that the parameter region 
$n>1$, $t_2>0$ is exactly equivalent to the region $n<1$, 
$t_2<0$ under the particle-hole transformation.  
By the same token, the region $n<1$, $t_2>0$ equals to 
the region $n>1$, $t_2<0$.  We therefore consider only 
the region $0<n\le 1$ for both positive and negative values 
of $t_2$.  

\section{METHOD}

The low-energy properties of TLL are characterized by a 
few quantities; most notably, the TLL parameter $K_\rho$ 
determines the long-range behavior of various correlation 
functions in the metallic TLL ground state.  It has however 
been recognized that the numerical calculation of $K_\rho$ 
for an arbitrary strength of correlations is very difficult.  
Recently, one of the authors has succeeded to overcome 
this difficulty,~\cite{Eji05} where a simple and stable method 
for calculating $K_\rho$ with the DMRG technique in single-band 
1D systems is proposed.  
In this section, we extend the method to the two-band systems 
and check the performance of the method by comparing the results 
with those obtained by the HF approximation which is known 
to provide a good estimation of $K_\rho$ in the weak-coupling 
regime.

\subsection{HF approximation}

It is known that the small-$U$ perturbative estimation of 
$K_\rho$ is feasible for $U \lesssim W/2$ in the 1D Hubbard 
model.~\cite{Sch90}  One of the authors applied this perturbative 
method to a two-leg Hubbard ladder model with four Fermi points 
and confirmed that it gives quantitatively reliable results in 
the weak-coupling regime ($U \lesssim W/4$).~\cite{San00} 
Since the low-energy physics of the two-leg ladders is 
equivalent to that of our model defined by 
Eq.~(\ref{hamiltonian}),~\cite{Fab96} we may naturally 
expect the perturbative estimation to be applicable to our 
case.  

In the TLL theory, the critical exponent $K_\rho$ is given by
\begin{equation}
K_\rho=\frac{1}{2}\sqrt{\pi \chi D},
\label{krho}
\end{equation}
where $\chi$ is the charge susceptibility defined as 
\begin{equation}
\chi^{-1}=\frac{1}{L}\frac{\partial^2 E_{\rm gs}(n)}{\partial n^2},
\label{chi}
\end{equation}
and $D$ is the Drude weight defined as 
\begin{equation}
D=\frac{\pi}{L}\frac{\partial^2 E_{\rm gs}(\phi)}{\partial \phi^2}.  
\label{d}
\end{equation}
Here, $L$ is the number of lattice site, $n$ is the band filling, 
$E_{\rm gs}$ is the ground-state energy, and $\phi$ is the magnetic 
flux.~\cite{Voi95}  Within the first-order perturbation expansion, 
$E_{\rm gs}$ can be determined as 
\begin{equation}
E_{\rm gs}=E_0+\frac{UL}{4}n^2
\label{energyperturbation}
\end{equation}
where $E_0$ is the ground-state energy of the corresponding 
noninteracting system.  We then obtain 
\begin{equation}
\chi^{-1}=\chi_0^{-1}+\frac{U}{2},~~~ D=D_0=\frac{4\chi_0^{-1}}{\pi}
\label{1stchiD}
\end{equation}
where $\chi_0$ and $D_0$ are the charge susceptibility and 
Drude weight of the noninteracting system, respectively.  
A simple expression for $K_\rho$ is therefore obtained as 
\begin{equation}
K_\rho \simeq \sqrt{\frac{2}{2+U\chi_0}}.
\label{1stperturbation}
\end{equation}
Note that this scheme is equivalent to the HF 
approximation.~\cite{San00}

\subsection{DMRG}

\begin{figure}[t]
\includegraphics[width= 6.5cm]{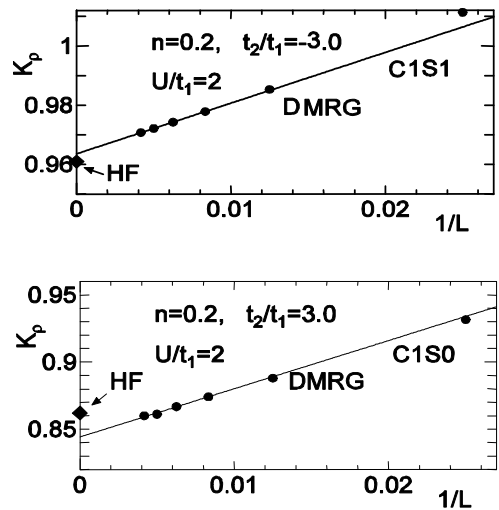}
\caption{Values of $K_\rho(L)$ calculated by the DMRG method 
and plotted as a function of the inverse system size $1/L$.  
Solid lines are the polynomial fits to the data for 
finite-size scaling analysis.  
Diamonds indicate the results calculated from the 
Hartree-Fock approximation.
The upper and lower panels show the result at $t_2/t_1=-3$ 
and $t_2/t_1=3$, respectively.  
$U/t_1=2$ and $n=0.2$ are assumed in both panels.}
\label{fig2}
\end{figure}

With the DMRG method, the TLL parameter $K_\rho$ is most 
generally obtained from the long-range decay of the density-density 
correlation.  The density-density correlation function is 
defined by the ground-state expectation value
\begin{equation}
C^{\rm NN}(r) 
= \frac{1}{L} \sum_{l=1}^{L} 
(\langle n_{l+r} n_l \rangle
- \langle n_{l+r} \rangle 
\langle n_l \rangle).
\label{CNNr}
\end{equation}
When the system has two Fermi points, it is known 
that the asymptotic behavior is given by
\begin{equation}
C^{\rm NN}(r) \sim
-\frac{K_{\rho}}{(\pi r)^2}
+\frac{A\cos(2k_{\rm F}r)}{r^{1+K_{\rho}}}\ln^{-3/2}(r)
+\cdots,
\label{eqn:den1}			 
\end{equation}
where $A$~is a constant.~\cite{Fra90,Sch90}  
We can extract $K_\rho$ via the Fourier transformation 
of Eq.~(\ref{CNNr}), 
\begin{equation}
C^{\rm NN}(q) 
= \frac{1}{L} \sum_{l=1}^{L} 
e^{-iqr} C^{\rm NN}(r),
\label{CNNq}
\end{equation}
where $0 \le q < 2\pi$.  
From the derivative at $q=0$, one finds the expression 
\begin{eqnarray}
K_{\rho}(L) = \frac{L}{2} C^{\rm NN} \left(\frac{2\pi}{L}\right),~~~ 
K_\rho = \lim_{L \to \infty} K_\rho(L). 
\label{Krho1}
\end{eqnarray}
for the thermodynamic limit.  
It has been demonstrated that the value of $K_\rho$ can 
be determined quite accurately by using Eq.~(\ref{Krho1}) 
with the DMRG method for the single-band Hubbard model.  
Thus, for the precise estimation, we need to calculate 
the density-density correlation function directly in the 
Fourier space; see Ref.~\onlinecite{Eji05} for further 
details. 

Let us now apply this scheme of estimating $K_\rho$ 
to a system with four Fermi points.  We then have to 
assume the asymptotic behavior of the density-density 
correlation function.  Here we assume the behavior 
\begin{equation}
C^{\rm NN}(r) \sim
-\frac{2K_{\rho}}{(\pi r)^2}
+\frac{B\cos[2(k_{\rm F1}-k_{\rm F2})r]}{r^{2k_\rho}}
+\cdots\; ,
\label{eqn:den2}			 
\end{equation}
in analogy with the case of two coupled chains,~\cite{Sch96}  
because the calculated low-energy excitation spectra 
of our model are similar to those of the two coupled 
chains.~\cite{Nis07}  We thus obtain
\begin{eqnarray}
K_{\rho}(L) =  \frac{L}{4} C^{\rm NN} \left(\frac{2\pi}{L}\right),~~~ 
K_\rho = \lim_{L \to \infty} K_\rho(L),
\label{Krho2}
\end{eqnarray}
as a substitute for Eq.~(\ref{Krho1}).  
In principle, one may calculate Eq.~(\ref{Krho2}) 
in the same way as Eq.~(\ref{Krho1}).  
However, the discarded weight in the DMRG 
calculation increases rapidly with increasing 
$|t_2/t_1|$, so that the calculation 
must be carried out more carefully.  

In this paper, we apply the open-end boundary conditions 
for precise DMRG calculations.~\cite{Whi92}  
We keep up to $m \simeq 4500$ density-matrix eigenstates 
in the DMRG procedure and extrapolate the calculated 
quantities to the limit $m \to \infty$.  We also use several 
chains with lengths $L=40$ to $240$ and then perform the 
finite-size scaling analysis based on the size-dependence 
of the quantities.  In this way, we can obtain the quite 
accurate ground state with an accuracy of 
$\Delta E_{\rm gs}/L \sim 10^{-6}-10^{-5}t_1$.  
In Fig.~\ref{fig2}, we demonstrate the finite-size scaling 
analysis for (a) the two- and (b) four-Fermi point cases.  
For both cases, one can see the systematic extrapolation 
of $K_\rho$ to the thermodynamic limit $L \to \infty$.  
We also find that, at least in this coupling strength 
$U/t_1=2$, a good agreement is obtained between the 
extrapolated values of $K_\rho$ obtained from the DMRG data 
and the corresponding values of $K_\rho$ obtained from 
the HF approximation.  

\section{RESULTS}

\subsection{Weak-coupling limit}

\begin{figure}[t]
\includegraphics[width= 7.5cm]{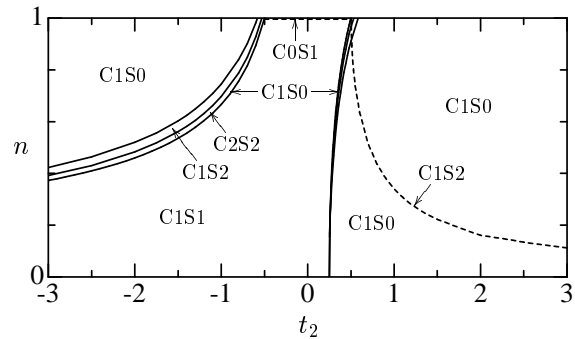}
\caption{Phase diagram of our model in the weak-coupling 
limit $U=0^+$.  We follow the notation of the symbol 
${\rm C}\alpha{\rm S}\beta$ of Ref.~\cite{Bal96}.}
\label{fig3}
\end{figure}

Let us first consider the phase diagram within a 
weak-coupling analysis.  Balents and Fisher 
have obtained the weak-coupling phase diagram 
of the two-leg Hubbard ladder model using the 
renormalization group technique and bosonization 
method.~\cite{Bal96}  Their analysis can be 
universally applied to a system with four Fermi 
points.  In addition, Fabrizio pointed out~\cite{Fab96} 
that the low-energy physics of the two-leg Hubbard 
ladder model is exactly the same as that of our 
model (\ref{hamiltonian}) via a simple mapping of 
the Fermi points, i.e., $\pm k_{\rm F1} \to \mp k_{\rm F}^a$ 
and $\pm k_{\rm F2} \to \mp k_{\rm F}^b$ where 
$\pm k_{\rm F}^a$ ($\pm k_{\rm F}^b$) are 
the Fermi points for the antibonding (bonding) band 
of the two-leg Hubbard ladder model.  
Then, Daul and Noack adapted the weak-coupling 
analysis of Ref.~\onlinecite{Bal96} for the analysis 
of the Hamiltonian (\ref{hamiltonian}).~\cite{Dau98}

In Fig.~\ref{fig3}, we show the phase diagram for 
$|t_2/t_1| < 3$ and $n<1$ in the weak-coupling 
limit $U=0^+$.  A notation ${\rm C}\alpha{\rm S}\beta$ 
denotes a phase with $\alpha$ gapless charge modes and 
$\beta$ gapless spin modes, where $\alpha$ and $\beta$ 
are integer values from $0$ to $2$.  Generally speaking, 
a metallic phase with four (two) Fermi points is 
characterized by ${\rm C}1{\rm S}0$ (${\rm C}1{\rm S}1$).  
Also, a spin-gapped liquid phase ${\rm C}1{\rm S}0$ 
appears around the FP critical boundary due to the van Hove 
singularity of the model.  
Note that the TLL parameter has the value $K_\rho = 1$ 
in nearly all the metallic regimes, except at the FP 
critical boundary and on the ${\rm C}0{\rm S}1$ line, 
where we have $K_\rho=1/2$.

\subsection{Small $U$}

\begin{figure}[t]
\includegraphics[width= 7.5cm]{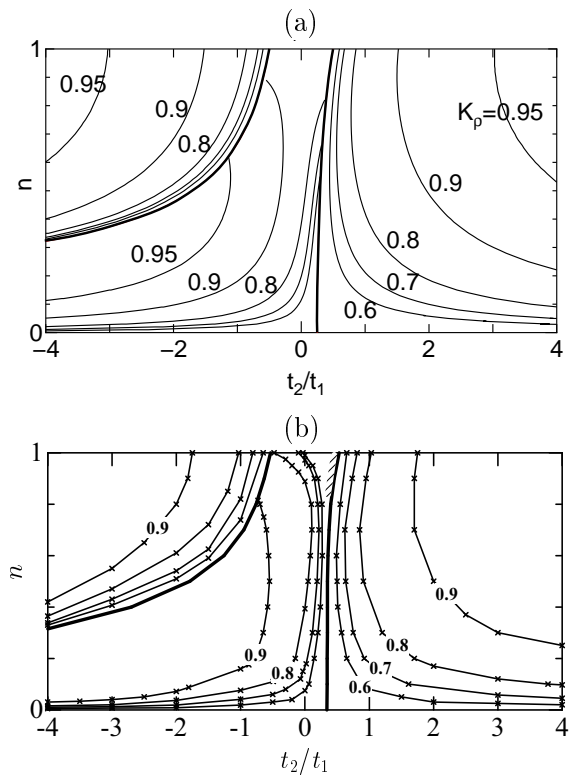}
\caption{Contour map of the TLL parameter $K_\rho$ in 
the $n$$-$$t_2/t_1$ plane for the weak-coupling 
interaction $U=2t_1$.  The results are obtained from 
(a) the HF and (b) DMRG calculations.  The thick line 
represents the FP critical boundary.}
\label{fig4}
\end{figure}

Let us turn to the small-$U$ perturbative regime where 
we choose the coupling strength $U=2t_1$.  In this regime, 
the HF estimation of $K_\rho$ [Eq.(\ref{1stperturbation})] 
is expected to give a good approximation.  When the system 
has two Fermi points, we can simply obtain 
\begin{eqnarray}
\chi_0^{-1}=\frac{\pi}{2} v_{\rm F},
\label{chi0two}
\end{eqnarray}
which leads to 
\begin{eqnarray}
K_\rho=\sqrt{\frac{\pi v_{\rm F}}{\pi v_{\rm F}+U}}
\label{Krhotwo}
\end{eqnarray}
where the Fermi velocity is
\begin{eqnarray}
v_{\rm F}=2t_1 \sin (\pi n/2) -4t_2 \sin (\pi n).
\label{Fermiv2}
\end{eqnarray}
This expression (\ref{Krhotwo}) is universal to the TLL 
with two Fermi points. 
On the other hand, when the system has four Fermi points, 
we obtain 
\begin{eqnarray}
\chi_0^{-1}=\frac{\pi}{2}\frac{|v_{\rm F1}v_{\rm F2}|}{|v_{\rm F1}|+|{v_{\rm F2}}|}
\label{chi0four}
\end{eqnarray}
after some calculations, where $v_{\rm F1}$ ($v_{\rm F2}$) 
is the Fermi velocity at the momentum $k_{\rm F1}$ ($k_{\rm F2}$) 
given as 
\begin{eqnarray}
v_{\rm F(1,2)}=2t_1 \sin (k_{\rm F(1,2)}a) -4t_2 \sin (2k_{\rm F(1,2)}a).
\label{Fermiv4}
\end{eqnarray}
Substituting Eq.~(\ref{chi0four}) for Eq.~(\ref{1stperturbation}), 
we obtain
\begin{eqnarray}
K_\rho=\sqrt{\frac{\pi v_{\rm F}^\ast}{\pi v_{\rm F}^\ast+U}}
\label{Krhofour}
\end{eqnarray}
with using an effective Fermi velocity
\begin{eqnarray}
v_{\rm F}^\ast \equiv \frac{|v_{\rm F1}v_{\rm F2}|}{|v_{\rm F1}|+|{v_{\rm F2}}|}. 
\label{effctv}
\end{eqnarray}

In Fig.~\ref{fig4}, we show the contour maps of $K_\rho$ at 
$U=2t_1$ in the parameter space of $t_2/t_1$ and $n$, which 
are calculated with (a) the HF approximation and 
(b) the DMRG method. 
The thick line represents the FP critical boundary and 
the thin lines form the contour map of $K_\rho$.  
We find that the quantitative agreement between the two phase 
diagrams is pretty good, which means that the HF scheme is 
still valid for this interaction strength $U/t_1=2$.  
In the entire region of the phase diagram, we find $K_\rho>1/2$ 
and thus the ground state may be described as the TLL.  
However, the deviation of the HF data from the DMRG data is 
relatively large around $t_2 \approx 0$ and $n \approx 1$, 
where the umklapp scattering becomes dominant. 

When the system has two Fermi points, the ground state can 
basically be presumed to be a standard 1D TLL.  
In both contour maps, $K_\rho$ becomes larger as 
$t_2$ decreases for fixed $n$.  This is because the 
effective interaction parameter $U/v_{\rm F}$ is reduced 
with decreasing $t_2$.  In particular, $K_\rho$ varies 
drastically around $t_2 \sim 0$ except when 
$n \lesssim 0.1$ and $n \gtrsim 0.9$, which is due to the 
rapid change of the inverse Fermi velocity $v_{\rm F}^{-1}$; 
for example, the Fermi velocity is estimated as 
$v_{\rm F}=\sqrt{2}t_1-4t_2$ at quarter filling ($n=1/2$).  
Note that $K_\rho=1/2$ is reached in the limits $n \to 0$ 
and $1$ and that $K_\rho>1/2$ everywhere else.

As soon as the Fermi surface splits from two into four 
points, $K_\rho$ drops (almost) discontinuously to $1/2$.  
When the system has four Fermi points, $v_{\rm F}^\ast$ goes 
to zero in the limits $v_{\rm F1} \to 0$ ($k_{F1} \to 0$) 
or $v_{\rm F2} \to 0$ ($k_{F2} \to \pi$) 
[corresponding to the diverging density of states].  
Consequently, the effective interaction parameter diverges, 
$U/v_{\rm F}^\ast \to \infty$, and the strong-coupling 
value $K_\rho = 1/2$ is produced.  This behavior is 
similar to that of the 1D single-band Hubbard model in the 
limit of $n \to 0$.  Then, $K_\rho$ increases rapidly as the 
parameters are away from the FP boundary line and gets closer 
to $1$ in the limit of $|t_2| \to \infty$.  In the two-band 
model, the criterion for the dominant superconducting correlation 
is $K_\rho > 1/2$.  We thus find that the superconducting 
correlation is the most dominant in the entire region with 
four Fermi points.

\subsection{Large $U$}

\begin{figure}[t]
\includegraphics[width= 7.5cm]{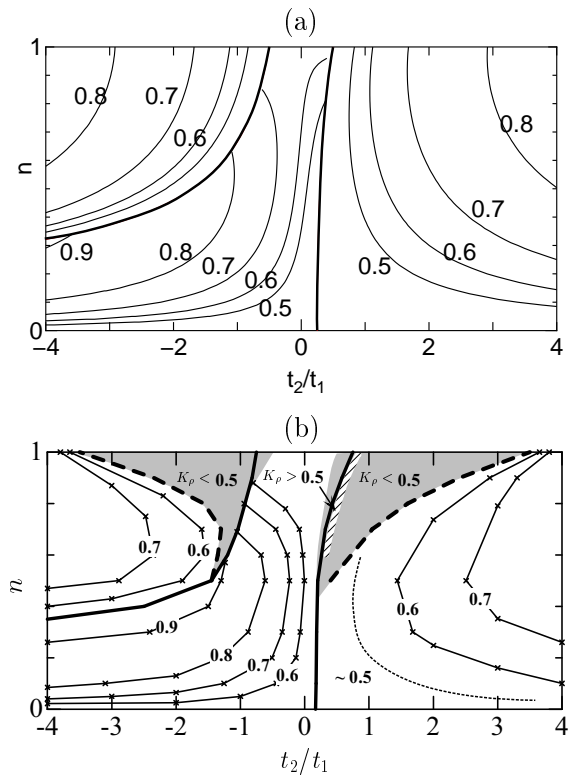}
\caption{Contour map of the TLL parameter $K_\rho$ in 
the $n$$-$$t_2/t_1$ plane for the strong-coupling 
interaction $U=10t_1$.  The results are obtained 
from (a) the HF and (b) DMRG calculations.  
The thick line represents the FP critical boundary.}
\label{fig5}
\end{figure}

Let us now consider how the small-$U$ contour map is 
affected by increasing the strength of the on-site 
Coulomb interaction.  
For large enough coupling $U$, it has been found that 
there is an extensive ferromagnetic 
phase~\cite{Pie96,Dau971,Dau972} and that the 
spin-triplet superconducting state is stabilized 
at the intermediate filling $n \sim 0.5$ when the 
next-nearest-neighbor hopping is large enough, 
$t_2/t_1 \gtrsim 2$.~\cite{Oht05}  
A breakdown of the TLL state was also reported for 
the 1D Hubbard and $t$$-$$J$ models with the 
next-nearest-neighbor hopping integrals,~\cite{Gro05,Ede97}  
where the latter model is essentially the same as our model 
(\ref{hamiltonian}) in the strong-coupling regime.

We here study the case of $U=10t_1$ as a typical 
interaction strength for realistic strongly-correlated 
electron systems.  In Fig.~\ref{fig5}, we show the contour 
map of $K_\rho$ obtained by (a) the HF approximation and 
(b) DMRG method.  We find that, at low densities 
($n \lesssim 0.4$), the agreement between the two 
contour maps is qualitatively good, while at intermediate 
to high densities ($n \gtrsim 0.4$), the situation 
seems to be totally different.  
In fact, the HF scheme is no longer appropriate and 
therefore the spin and charge fluctuations have to be 
taken seriously into account beyond the usual weak-coupling 
picture. 
We will thus proceed to a discussion based on the DMRG 
contour map in the following.  

We first note that, as far as the system has two Fermi 
points, the basic properties are qualitatively the same as 
those of the weak-coupling regime.  Thus, the behavior 
of $K_\rho$ at $U=10t_1$ is still similar to that at 
$U=2t_1$ although the value of $K_\rho$ becomes relatively 
small.  We also note that the FP boundary is (slightly) 
shifted toward the smaller $|t_2|$ direction due to 
renormalization of the band structure at $U>0$, as was 
pointed out in Ref. \onlinecite{Dau98}.  At the same time, 
the FP boundary line is somewhat blurred because of some 
strong quantum effects; the change in $K_\rho$ at the FP 
boundary is still sharp but no longer discontinuous as 
in the weak-coupling limit. 

Let us turn to the case with four Fermi points.  
The ground state is affected drastically by the (strong) 
interaction strength.  Unlike in the small-$U$ contour map, 
we find that there is a substantial region with $K_\rho<0.5$ 
around half filling [denoted by the shadowed area 
in Fig.~\ref{fig5}(b)].  According to the TLL theory, 
this value of $K_\rho$ is possible only when long-range 
repulsive interactions are included in the model.  Therefore, 
the ground state in the shadowed regime would no longer 
belong to the general class of the TLLs.  This is consistent 
with the previous DMRG results.~\cite{Dau98,Gro05}  
As discussed in the next subsection, this non-TLL-like regime 
consists of a spin-gapped phase and a paramagnetic phase with strong 
ferromagnetic fluctuations.  Also, it is interesting to note 
that the TLL parameter remains constant $K_\rho \sim 0.5$ in 
a wide region of the phase diagram near the FP boundary 
at low densities.  

Furthermore, it is particularly worth noting that $K_\rho$ 
seems to be enhanced significantly at the FP boundary near 
half filling.  [In this area, precise evaluation of $K_\rho$ 
in the thermodynamic limit $1/L \to 0$ is rather hard 
because $|\partial K_\rho(L)/\partial(1/L)|$ increases with 
decreasing the inverse system size $1/L$.]  
If the definition (\ref{krho}) could be still valid in this 
region, $K_\rho$ become quite large $>1$: for the van Hove 
singularity, the charge susceptibility $\chi$ in Eq.~(\ref{chi}) 
can diverge and the Drude weight $D$ in Eq.~(\ref{d}) must 
remain finite.  
This may be associated with the ${\rm C}1{\rm S}0$ phase 
attributed to the van Hove singularity at $k_{F1}=0$ 
(or $k_{F2}=\pi$) in the weak-coupling phase diagram.  
This is consistent with a prediction that the 
superconducting fluctuations increase with increasing 
the difference between $|v_{\rm F1}|$ and $|v_{\rm F2}|$, 
as was suggested in Ref.~\onlinecite{Bal96}.  

\subsection{Spin gap}

\begin{figure}[t]
\includegraphics[width= 8.0cm]{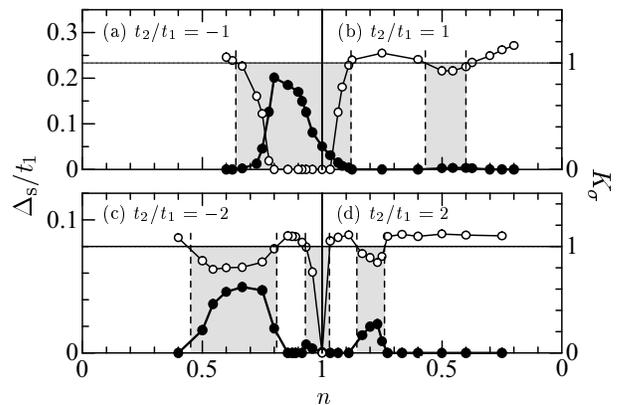}
\caption{Calculated values of the spin gap (solid symbols, 
left axis) and TLL spin exponent (open symbols, right axis) 
at (a) $t_2/t_1=-1$, (b) $1$, (c) $-2$, and (d) $2$.  
$U/t_1=10$ is assumed. 
The solid line denotes $K_\sigma=1$ and the vertical 
dotted lines indicate the critical boundaries between 
the presence and absence of the spin gap.}
\label{fig6}
\end{figure}

For more elaborate studies of the region with four Fermi 
points, we consider the spin degrees of freedom in 
the strong-coupling regime $U=10t_1$.  
Of particular interest here is the presence or absence 
of a finite energy gap in the spin excitation spectrum.  
We thus evaluate the spin gap defined by an energy 
difference between the first spin-triplet excited 
state and the singlet ground state: i.e., 
\begin{equation}
\Delta_{\rm s}=\lim_{L \to \infty}[E_{\rm gs}(N,1)-E_{\rm gs}(N,0)], 
\end{equation}
where $E_{\rm gs}(N,S_{\rm z})$ is the ground-state energy for a 
given number of electrons $N$ and z-component of the 
total spin $S_{\rm z}$.  It is however known that, for 
some parameter values, the spin gap becomes too small 
to figure out if it remains finite, e.g., 
$\Delta_{\rm s} \lesssim 10^{-3}t_1$.  
For verifying the presence or absence of the spin gap, 
we then calculate the TLL spin exponents, which is given 
by
\begin{equation}
K_\sigma=\lim_{L \to \infty}\frac{L}{2}\sum_{kl}
e^{i\frac{2\pi}{L}(k-l)}\left\langle S^z_k S^z_l \right\rangle
\end{equation}
where $S^z_i=n_{i\uparrow}-n_{i\downarrow}$.  
We should find that the spin exponent takes the value 
$K_\sigma=0$ in the spin-gapped phase and $K_\sigma=1$ 
everywhere else in the thermodynamic limit.~\cite{Voi92} 
However, for finite-size systems, the situation is 
not so simple.  In practice, in the spin-gapless 
phase, one cannot expect to find the value $K_\sigma \to 1$ 
due to the logarithmic corrections.  It is known 
that the logarithmic corrections vanish at which 
the spin gap opens, in analogy with the dimerization 
transition in the $J_1$-$J_2$ model 
[see Eq.~(\ref{heisenberg}) below].~\cite{Egg96}  
Also, in the spin-gapped phase, if the spin gap is very 
small, the convergence of $K_\sigma$ to $0$ will obviously 
occur only for very large systems.  Therefore, we here 
determine the critical point at which the spin gap opens 
by adopting the condition that the value of 
$K_\rho$ crosses $1$.  
This method was first proven to be useful in Ref.~\cite{Sen02}  
In Fig.~\ref{fig6}(a)-(d), the spin gap and TLL spin 
exponent calculated by the DMRG method are plotted as 
a function of the band filling $n$ for various $t_2/t_1$ 
values.  

For $U\gg t_1$, $t_2$, our model (\ref{hamiltonian}) 
at half filling can be mapped onto a Heisenberg model 
\begin{eqnarray}
H=J_1 \sum_i \vec{S}_i\cdot\vec{S}_{i+1}
+J_2\sum_i\vec{S}_i\cdot\vec{S}_{i+2}
\label{heisenberg}
\end{eqnarray}
with $J_1=4t_1^2/U$ and $J_1=4t_2^2/U$.  This model has 
been extensively studied both 
analytically~\cite{Hal82,Oka92,Egg96,Whi96} and 
numerically.~\cite{Ari98,Dau00}  
It has been found that the spin gap opens when 
$J_2/J_1 \ge 0.241$; the ground state is of a dimerized 
zigzag-bond state for $0.241 \le J_2/J_1 \lesssim 0.5$ 
and of the Majumdar-Ghosh state with incommensurate 
spiral correlations for $J_2/J_1 \gtrsim 0.5$.  
For example, the spin gap was estimated to be 
$\Delta_{\rm s}\simeq 0.25J_1$ at $J_2/J_1\sim(t_2/t_1)^2=1$ 
in the previous DMRG study.~\cite{Whi96}  This value is 
comparable to our estimation $\Delta_{\rm s}/t_1\simeq 0.05$ 
for $t_2/t_1=1$ and $U=10t_1$.  
Also, the spin gap is of an exponential dependence on $J_2/J_1$ 
as $\Delta_{\rm s}\propto\exp\big(-{\rm const.}\times J_2/J_1\big)$ 
for large $J_2/J_1$ values, which is consistent with a very 
small spin gap $\Delta_{\rm s}/t_1\sim 0.0005$ for 
$t_2/t_1=2$ and $U=10t_1$ obtained in our calculations.  

Let us turn to the evolution of the spin gap upon doping.  
In the weak-coupling phase diagram, a metallic phase with 
four (two) Fermi points is simply characterized as the 
spin-gapped (gapless) TLL.  However, for large $U$ values, 
it is difficult to speculate the $n$-dependence of the 
spin gap because of the competition between the 
antiferromagnetic exchange interaction and two kinds of 
the ferromagnetic interactions;  One is induced by the 
Nagaoka mechanism, which leads to long-range ferromagnetic 
fluctuations for slightly-doped systems,~\cite{Nag66,Mat74} 
where the mechanism is known to work even for finite 
doping levels.~\cite{Mue93,Geb91} 
The other is the three-site ring-exchange mechanism, 
which yields ferromagnetic spin correlations for the 
intermediate filling.~\cite{Oht05}  This mechanism works 
only when the product of three hopping integrals along 
the triangles forming the triangular lattice is positive, 
i.e., $t_1^2t_2>0$ in our system [see, Fig.~\ref{fig1}(a)].  
Away from half filling, the spin gap has so far been 
calculated with the DMRG methods for some 
parameters,~\cite{Dau98,Oht05,Oku07} and we now 
study the spin gap in a wider range of 
the $n$$-$$t_2/t_1$ plane.  
The results are the following: 

(i) For $t_2/t_1=-1$, $n<1$ [Fig.~\ref{fig6}(a)], 
the spin gap is considerably enhanced with decreasing 
$n$ near half filling.  Since the geometrical spin 
frustration is reduced by doping, the spin-singlet bound 
state is in a better position to be formed.  
The value of $\Delta_{\rm s}$ increases rapidly 
as $n$ decreases, reaches the maximum value $\Delta_{\rm s}/t_1\sim 0.2$ 
around $n \sim 0.8$, and goes down to zero at the FP boundary 
$n \sim 0.64$.  The spin gap is always zero when the system 
has two Fermi points, which is in agreement with the 
weak-coupling phase diagram.  
We should note that no singularity in the spin gap is found 
at half filling.

(ii) For $t_2/t_1=-2$, $n<1$ [Fig.~\ref{fig6}(c)], 
the spin gap is enhanced by doping in the vicinity 
of half filling as in the case of (i).  However, unlike in 
the case (i), the spin gap vanishes around $n \sim 0.93$ 
even though the system has still four Fermi points.  
This may be related to the ferromagnetic spin fluctuations 
induced by the Nagaoka mechanism.  Thus, the region for 
$0.81 \lesssim n \lesssim 0.93$ is spin gapless.  
Further away from half filling, the Nagaoka mechanism 
can no longer work well and the ferromagnetic fluctuations 
weaken.  Consequently, the spin gap opens again in the 
filling from $n \sim 0.81$ to the FP critical boundary 
$n \sim 0.45$.  
It is also interesting to note that the critical filling 
$n \sim 0.81$ agrees with the TLL critical boundary; 
i.e., the spin gap starts to open at the point of $K_\rho=0.5$. 

(iii) For $t_2/t_1=1$, $n<1$ [Fig.~\ref{fig6}(b)], 
the spin gap is always finite for small $U$ values 
[see Ref.\onlinecite{Dau98}].  For large $U$ values, however, 
the spin gap behaves intricately as a function of doping 
due to the existence of two types of the ferromagnetic 
fluctuations. 
Near half filling, $\Delta_{\rm s}$ decreases with 
decreasing $n$ and disappears around $n \sim 0.88$.  
We can see that no spin gap exists in a region 
$0.57 \lesssim n \lesssim 0.88$.  
This is consistent with the fact that the critical 
interaction strength $U_c$ at the ferromagnetic transition 
is relatively small for this region,~\cite{Dau98} which 
would indicate the strong ferromagnetic fluctuations there.  
And then, with decreasing $n$, the spin gap opens again 
around $n \sim 0.57$, where the Nagaoka mechanism no 
longer works well.  Like in the case (ii), the point 
where the spin gap opens is on the TLL critical boundary.  
By further doping, the spin gap closes around $n \sim 0.4$.  
For $n \lesssim 0.4$, the spin gap is zero due to the 
ferromagnetic fluctuations induced by the three-site 
ring-exchange interaction.  Note that there is no spin gapless 
region derived by the three-site ring-exchange interaction 
in the case of (ii) where $t_1^2t_2<0$.

(iv) For $t_2/t_1=2$, $n<1$ [Fig.~\ref{fig6}(d)], 
the properties are qualitatively the same as in the case of (iii).  
The spin gap remains finite only for a tiny region in the vicinity 
of half filling ($n \gtrsim 0.97$) and for a small region 
adjacent to the TLL critical boundary 
($0.74 \lesssim n \lesssim 0.85$).  The three-site 
ring-exchange interaction is more robust 
than in the case of (iii),~\cite{Oht05} 
so that the gapless area becomes wider ($n \lesssim 0.74$). 

\subsection{Fully polarized state}

\begin{figure}[t]
\includegraphics[width= 6.5cm]{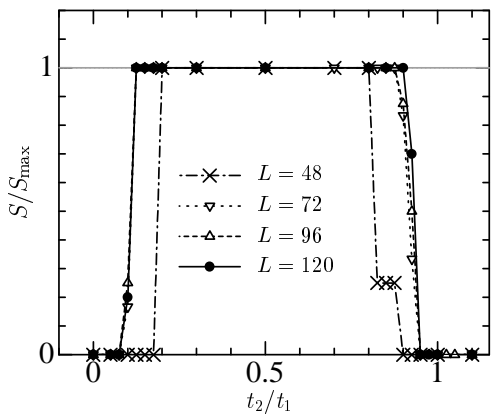}
\caption{Calculated values of the total-spin quantum 
number $S$ as a function of $t_2/t_1$ for various 
system sizes $L$.  $U/t_1=10$ and $n=1/6$ are assumed.}
\label{fig7}
\end{figure}

Of further interest is the presence of the fully polarized 
ferromagnetic state, which occurs when $t_2$ is positive 
in the strong-coupling regime.  Previously, for $U=\infty$, 
ferromagnetism has been analytically shown to exist in the 
three limiting cases: $n \to 1$,~\cite{Mat74} 
$t_2 \to 0$,~\cite{Sig92} and $n \to 0$.~\cite{Mue93}.  
Also, for finite $U$ values, it has been shown numerically that 
there is an extensive ferromagnetic phase, where the exact 
diagonalization,~\cite{Pie96} variational,~\cite{Dau971} 
and DMRG~\cite{Dau972} methods have been used.  

Let us then investigate how the ferromagnetic phase appears 
in the phase diagram.  
We can find it by calculating the expectation value of total-spin 
operator $\vec{S}$ in the ground state, which is defined by
\begin{equation}
\left\langle\vec{S}^2\right\rangle=
\sum_{ij}\left\langle\vec{S}_i\cdot\vec{S}_j\right\rangle
=S(S+1).
\end{equation}
For a fully-polarized state, one will obtain $S=S_{\rm max}=N/2$, 
i.e., $S/S_{\rm max}=1$.  In Fig.~\ref{fig7}, we show the 
total spin $S$ normalized with respect to $S_{\rm max}$ 
as a function of $t_2/t_1$ at $U=10$ and $n=1/6$ for various 
system sizes.  We can see a transition between paramagnetic 
and ferromagnetic states at $(t_2/t_1)_{\rm c} \sim 0.1$ 
and $0.95$.  The change in $S/S_{\rm max}$ at $(t_2/t_1)_{\rm c}$ 
becomes sharper with increasing $L$, suggesting the transition 
to be (almost) discontinuous for $L>72$.  Therefore, the 
transition may be of the first-order in the thermodynamic limit.  
Thus, the critical transition point can be determined in the 
parameter space, which will be given in the next subsection.  

\subsection{Phase diagram}

\begin{figure}[t]
\includegraphics[width= 7.5cm]{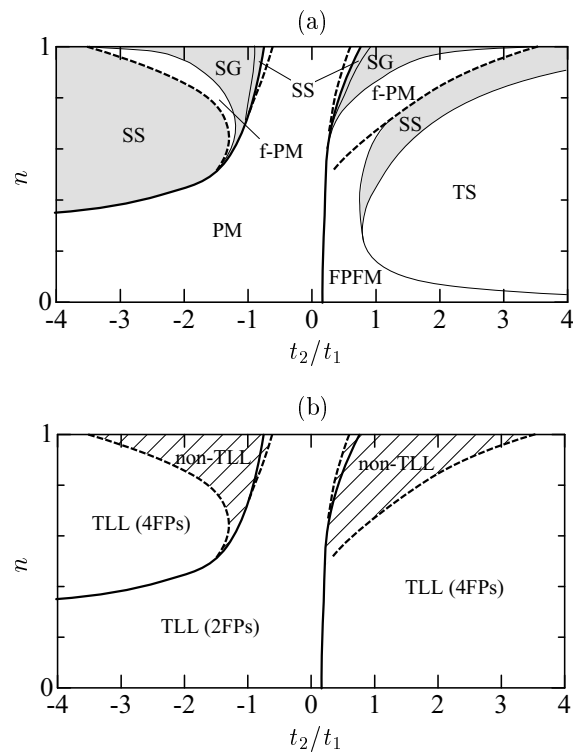}
\caption{
(a) Ground-state phase diagram of the 1D Hubbard model 
with the next-nearest-neighbor hopping, calculated by 
the DMRG method.  $U/t_1=10$ is assumed.  
The bold (dotted) line indicates the FP (TLL) critical boundary 
and the shadowed area represents a spin-gapped phase.  
We use the following abbreviations; 
PM: paramagnetic metal, 
SG: spin-gapped liquid with incommensurate spiral correlations, 
f-PM: paramagnetic metal with strong ferromagnetic fluctuations, 
FPFM: fully polarized ferromagnetic metal, and 
SS (TS): spin-singlet (triplet) superconductivity. 
(b) Boundary lines between the TLL and non-TLL-like regions.}
\label{fig8}
\end{figure}

Based on the calculated results of the TLL parameters $K_{\rho,\sigma}$, 
spin gap $\Delta_{\rm s}$, and total-spin quantum number $S$, 
we draw a $U=10$ phase diagram of the 1D Hubbard model with the 
next-nearest-neighbor hopping.  The result is shown in Fig.~\ref{fig8}(a).  
We find that our system Eq.~(\ref{hamiltonian}) exhibits a variety 
of phases in the parameter space of $t_2/t_1$ and $n$; it includes 
a paramagnetic metal (PM), 
a spin-gapped liquid with incommensurate spiral correlations (SG), 
a paramagnetic metal with strong ferromagnetic fluctuations (f-PM), 
a fully-polarized ferromagnetic metal (FPFM), 
a spin-singlet superconductivity (SS), and 
a spin-triplet superconductivity (TS).  
The bold and dotted lines in Fig.~\ref{fig8}(a) indicate the FP 
and TLL critical boundary, respectively, and the shadowed area 
represents a spin-gapped phase. 

When the system has two Fermi points, the ground-state properties 
are essentially the same as those of the standard 1D Hubbard model 
with $t_2=0$.  The system is thus a paramagnetic metal with 
$0.5 < K_\rho < 1$, where the $2k_{\rm F}$-SDW correlation is 
most dominant.  
The introduction of $t_2$ brings a sort of frustration to the 
$2k_{\rm F}$-SDW oscillation, but the oscillation is never broken 
down as long as the system has two Fermi points.

We now turn to the region with four Fermi points.  
As shown in Sec.~IV.~D, this region consists of the TLL 
($K_\rho \ge 1/2$) and non-TLL-like ($K_\rho<1/2$) phases 
[see Fig.~\ref{fig8}(b)].  
Looking first at the TLL phase with $K_\rho > 1/2$, the 
superconducting correlation is most dominant according 
to the TLL theory.  
The superconducting phase for $t_2>0$ is further divided 
into a couple of phases, depending on the presence or absence 
of the spin gap; the spin-gapless phase extends over a wide 
range for large $t_2$, which is in contrast to the weak-coupling 
phase diagram.  This spin-gapless area seems to be expanded 
by increasing the on-site interaction $U$, as compared to 
the previous DMRG results for $U=2t_1$.~\cite{Dau00}  
[In Ref.\onlinecite{Dau00}, this phase is characterized as 
$2 \times {\rm C}1{\rm S}1 = {\rm C}2{\rm S}2$.]  
The ground state is featured as the spin-triplet superconductivity, 
as has been confirmed numerically.~\cite{Oht05}  
On the other hand, the spin-gapped phase is characterized 
by the spin-singlet superconductivity, which is remnant of 
a wide ${\rm C}1{\rm S}0$ region in the weak-coupling phase 
diagram.  
It is particularly worth noting that the TLL phase for $t_2<0$ 
is always spin-gapped and the spin-singlet superconducting 
correlation is most dominant.  This is consistent with 
the fact that the three-site ring-exchange interaction 
for spin-triplet coupling does not work if $t_1^2t_2<0$. 

The other TLL phase belongs to the fully-polarized 
ferromagnetic metal near the FP boundary and at low densities.  
A nearly flat-band system is realized since the two band 
minima are slightly occupied by electrons at low densities 
(or the band maximum at $k=0$ is slightly occupied by holes 
near the FP boundary).  
Consequently, the ferromagnetic ground state is stabilized.  
In the FPFM phase, we estimate the TLL parameter as 
$K_\rho \sim 0.5$ [see Fig.~\ref{fig5}(b)], which is the 
same as that of a spinless fermion system.

We next consider the non-TLL-like regime, which extends 
between the TLL region and the half-filling line.  
The paramagnetic phase with strong ferromagnetic correlation 
is located in the vicinity of the TLL regime, where the 
ferromagnetic fluctuations are enhanced due to the Nagaoka 
mechanism; this phase is denoted as f-PM in Fig.~\ref{fig7}.  
At present value of $U$, the total spin of the ground state 
is zero in the entire area of the f-PM phase.  As $U$ increases, 
the f-PM phase is enlarged and the system would be fully polarized 
at a critical value of $U$. 

Further approaching the vicinity of $n=1$, the spin-gapped 
region appears again.  Although most of the spin-gapped region 
is paramagnetic, a narrow spin-singlet superconducting 
phase with $K_\rho>0.5$ exists along the FP boundary line.  
We can interpret this situation by assuming the system to be 
a slightly doped $J_1$-$J_2$ Heisenberg model (\ref{heisenberg}): 
for $J_2/J_1 \gtrsim 0.5$, the ground state is of the 
Majumdar-Ghosh state with incommensurate spiral correlations 
and the spin-singlet bound state is formed along the $t_2$-chains, 
where the spin-singlet bound state cannot easily move.  
On the other hand, for $0.241 \le J_2/J_1 \lesssim 0.5$, 
the ground state is of a dimerized zigzag-bond state where 
the spin-singlet bound state is formed between the two 
$t_2$-chains.  At finite doping levels of holes, the spin-singlet 
pairs are mobile in this region, so that in the ground state 
an additional pair of holes is actually confined to a `rung' 
because the gain in kinetic energy due to the hole motion is 
larger than the combined loss in the pairing energy and 
kinetic energy of the spin dimers in the Majumdar-Ghosh state.  
Thus, the narrow spin-singlet superconducting state can be 
regarded as the doped zigzag-bond state.  

\section{SUMMARY}

We have studied the 1D Hubbard model with the 
nearest-neighbor and next-nearest-neighbor hopping integrals 
by using the DMRG method and HF approximation.  
Based on the calculated results of the TLL parameters, spin 
gap, and total-spin quantum number, we have determined the 
ground-state phase diagrams in the weak-coupling ($U=2t_1$) 
and strong-coupling ($U=10t_1$) regimes.  Surprisingly, the 
strong-coupling phase diagram contains a large variety of 
distinct phases, depending on the hopping integrals and band 
filling.  

We have found for $U=2t_1$ that the HF results agrees well 
with the DMRG results except for $t_2 \approx 0$ and $n \approx 1$ 
where the umklapp scattering strength is dominant.  
The phase diagram is qualitatively the same as that in 
the weak-coupling limit.
When the system has two Fermi points, the $2k_{\rm F}$-SDW 
correlation is most dominant with $1/2 < K_\rho < 1$.  
As soon as the Fermi surface splits from two into four points, 
the parameter $K_\rho$ drops (almost) discontinuously to $1/2$.  
We then have found $K_\rho > 1/2$ in the entire region with 
four Fermi points and thus the superconducting correlation is 
most dominant. 

We have then found for $U=10t_1$ that the HF results no longer 
agree with the DMRG results because the umklapp scattering 
strength becomes very large.  Due to the unconventional combination 
of the charge and spin degrees of freedom induced by the 
next-nearest-neighbor hopping, the system can accommodate a 
variety of physical phenomena unparalleled in the simple 1D 
Hubbard model.  
The region with two Fermi points is characterized by the 
$2k_{\rm F}$-SDW phase as in the case of $U=2t_1$.  
However, the region with four Fermi points is drastically 
affected by the Coulomb interaction and the breakdown of the 
TLL state occurs near half filling.  
The strong-coupling phase diagram contains a large number of 
distinct metallic phases; namely, a paramagnetic metal, 
a spin-gapped liquid with incommensurate spiral correlations, 
a paramagnetic metal with strong ferromagnetic fluctuations, 
a fully-polarized ferromagnetic metal, 
a spin-singlet superconductivity, and 
a spin-triplet superconductivity. 

In contrast to the weak-coupling phase diagram which predicts 
a spin-gapped liquid (or superconducting) phase when the system 
has four Fermi points, we have found that the spin gap vanishes 
in the substantial region in the strong-coupling phase diagram.   
The absence of the spin gap is derived by three types of 
ferromagnetic mechanisms.  The first is the flat-band mechanism 
around the FP boundary and at low densities.  
A nearly flat band is realized since the two band minima 
are slightly occupied by electrons at low densities 
(or the band maximum at $k=0$ is slightly occupied by holes 
near the FP boundary.)  Thus, the ground state is stabilized as 
the fully polarized ferromagnetic state.  
The second is the three-site ring-exchange interaction for 
$t_2 \gtrsim t_1$ at intermediate filling, where all the 
triangles formed by the hopping integrals satisfy the 
ferromagnetic sign rule $t_1^2t_2>0$.~\cite{Oht05}  
The ferromagnetic interaction is short ranged~\cite{Nis07} 
and is not sufficient to make the system ferromagnetic.  
The third is the Nagaoka mechanism near half filling.  
Since the competing antiferromagnetic exchange interaction is 
large, the ground state is not fully polarized.  

Concerning the superconductivity, we have found a couple of new 
phases, which are absent in the weak-coupling limit.  One is the 
spin-triplet superconducting phase for $t_2 \gtrsim t_1$, where 
the attractive interaction is caused by the gain in kinetic energy 
due to ring exchange of electrons.  The other is the spin-singlet 
superconducting phase along the FP boundary near half filling, 
where the superconducting fluctuations are enhanced by large 
difference between two Fermi velocities.  This state is also 
regarded as the doped zigzag-bond spin-gapped state.

\acknowledgments
This work was supported in part by Grants-in-Aid for Scientific 
Research (Nos. 18028008, 18043006, and 18540338) from the Ministry 
of Education, Science, Sports, and Culture of Japan. 
A part of computations was carried out at the Research Center 
for Computational Science, Okazaki Research Facilities, and 
the Institute for Solid State Physics, University of Tokyo.

\end{document}